\documentclass[conference,a4paper]{IEEEtran}
\IEEEoverridecommandlockouts
\usepackage{cite}
\usepackage{amsmath,amssymb,amsfonts}
\usepackage{graphicx}
\usepackage{textcomp}
\usepackage{xcolor}

\usepackage[linesnumbered, ruled]{algorithm2e}
\SetKwRepeat{Do}{do}{while}%

\pdfoutput=1

\usepackage{caption}

\usepackage{subcaption,graphicx}

\usepackage{tikz}
\newcommand*\circled[1]{\tikz[baseline=(char.base)]{
    \node[shape=circle,draw,inner sep=1pt] (char) {#1};}}

\newcommand*\squaredNum[1]{\tikz[baseline=(char.base)]{
    \node[shape=rectangle,draw,inner sep=1pt] (char) {#1};}}

\def\BibTeX{{\rm B\kern-.05em{\sc i\kern-.025em b}\kern-.08em
    T\kern-.1667em\lower.7ex\hbox{E}\kern-.125emX}}
\begin{document}

\title{Privacy-Friendly Peer-to-Peer Energy Trading: A Game Theoretical Approach \\
\thanks{This work was supported in part by the EPSRC through the projects EnnCore EP/T026995/1 and SCorCH EP/V000497/1 and by the Flemish Government through the FWO SBO project SNIPPET S007619. K.E is funded by The Ministry of National Education, Republic of Turkey. M.A.M. is funded by the DKO Fellowship awarded by The University of Manchester.}
}

\author{
    \IEEEauthorblockN{
        Kamil Erdayandi\IEEEauthorrefmark{1}, Amrit Paudel\IEEEauthorrefmark{3}, Lucas Cordeiro\IEEEauthorrefmark{1} and 
        Mustafa A. Mustafa\IEEEauthorrefmark{1}\IEEEauthorrefmark{2}}
\IEEEauthorblockA{
    \IEEEauthorrefmark{1}\textit{Department of Computer Science }, \textit{The University of Manchester}, UK\\
    \IEEEauthorrefmark{2}\textit{imec-COSIC}, \textit{KU Leuven}, Belgium \\
    \IEEEauthorrefmark{3}\textit{The University of British Columbia}, Canada \\
    Email:  \IEEEauthorrefmark{1}name.surname@manchester.ac.uk, 
    \IEEEauthorrefmark{3}amrit.paudel@ubc.ca
    }
    
}

\maketitle

\begin{abstract}
In this paper, we propose a decentralized, privacy-friendly energy trading platform (PFET) based on game theoretical approach -- specifically Stackelberg competition. Unlike existing trading schemes, PFET provides a competitive market in which prices and demands are determined based on competition, and computations are performed in a decentralized manner which does not rely on trusted third parties. It uses homomorphic encryption cryptosystem to encrypt sensitive information of buyers and sellers such as sellers' prices and buyers' demands. Buyers calculate total demand on particular seller using an encrypted data and sensitive buyer profile data is hidden from sellers. Hence, privacy of both sellers and buyers is preserved.  Through privacy analysis and performance evaluation, we show that PFET preserves users’ privacy in an efficient manner.

\end{abstract}


\section{Introduction}

Electricity generation 
 is slowly transitioning to Renewable Energy Sources (RES) such as wind and solar~\cite{BP}.
However, this transition brings new challenges. 
RES are not stable energy sources as their output fluctuates based on weather conditions~\cite{liang2016emerging}. This adds uncertainty to the generation side, in addition to the uncertainty on the demand side, making balancing the grid more challenging and less efficient. Unfortunately, traditional electricity markets offering two-tier (peak and off-peak) retail pricing for buyers and Feed-in-Tariffs (FiTs) for sellers are not effective enough to deal with these uncertainties.   

To address this issue, Peer-to-Peer (P2P) electricity trading markets have been proposed~\cite{paudel2018peer,morstyn2018bilateral,paudel9102274,lee2014direct,tushar2018peer}. 
They aim to incentivise users to be more proactive by allowing them to trade electricity between each other for more favourable prices than the retail prices and FiTs.   
 Hence, RES owners can collaboratively or individually maximise their profits and reduce their bills by trading electricity directly with other users. 

However, these trading markets require data sharing which may pose threats to privacy of users~\cite{Mustafa2016_sec_anal, Kalogridis2014}. 
For example, some entities may use other users’ offers and bids information to infer who is selling or buying how much electricity and when. In addition, prices offered by sellers can also reveal private data about energy usage pattern of seller prosumer. Such data is closely correlated to users’ consumption patterns. These situations may create privacy risks in which private information of the users may be leaked~\cite{montakhabi2020sharing}. 

In order to mitigate and alleviate these risks, use of various techniques have been proposed
~\cite{dimitriou2013privacy,radi2019privacy,li2018privacy,abidin2016mpc, Abidin2018, liu2020panda,xie2020privacy}. 
Dimitriou et al.~{\cite{dimitriou2013privacy}} and Radi et al.~{\cite{radi2019privacy}} use anonymisation techniques to hide users' identity. However, these can be reversed by using techniques described in~\cite{jawurek2011smart}. 
 A privacy-preserving prepayment-based energy trading platform is proposed in~\cite{li2018privacy}. However, the proposed platform has a centralised structure and is vulnerable to single point of failure. Privacy preserving double auction mechanisms are proposed in~\cite{abidin2016mpc,Abidin2018}~and~{\cite{liu2020panda}}, which use Multi-Party Computation (MPC) and Homomorphic Encryption (HE) schemes, respectively. Solely, only Xie at al.~\cite{xie2020privacy} proposed a game theoretical trading mechanism based on HE. However, the proposed market is not competitive. A \textit{fixed} market price is determined by the \textit{buyers}, and trading is performed over this price. Cooperation techniques to leverage better prices with group decisions are yet to be implemented.

{To address these limitations, we propose a novel Privacy-Friendly Electricity Trading (PFET) platform that provides a \textit{competitive} market for users based on a game-theoretical approach (Stackelberg Game) while protecting users' privacy by deploying Homomorphic Encryption (HE) scheme. To the best of our knowledge, this is the first \textit{competitive} game theoretical approach in energy trading systems that utilises HE and does not require TTPs. We implement and evaluate the performance of PFET to demonstrate it's effectiveness for communities with different number of buyers and sellers}. 

Paper organisation: Design preliminaries 
are given in Section~\ref{preliminaries}. Section~\ref{sec:PFET} presents our PFET.  Sections~\ref{equilibrium_time_complexity},~\ref{security_analysis}~and~\ref{performance_evaluation} evaluate PFET in terms of equilibrium and time complexity, privacy, and performance, respectively. Section~\ref{conlcusion_future_work} concludes the paper and gives directions for future work.

\section{Design Preliminaries}\label{preliminaries}


\subsection{System Model and Iterations}\label{system_model}

The P2P trading market used in our design consists of \textit{sellers} and \textit{buyers}. It is modelled as a Stacklelberg game in which sellers form a leader team, while buyers a follower team. In the first iteration, sellers make  decisions  and  undertake  strategies, and buyers follow them and respond back with their own proposal and strategies. In the following iterations, sellers update their strategies according to responses from buyers and buyers update theirs accordingly until the market reaches to a point where any further updates on strategies are not beneficial. 

\begin{figure}[!t]
\centering
\includegraphics[width=0.32\textwidth]{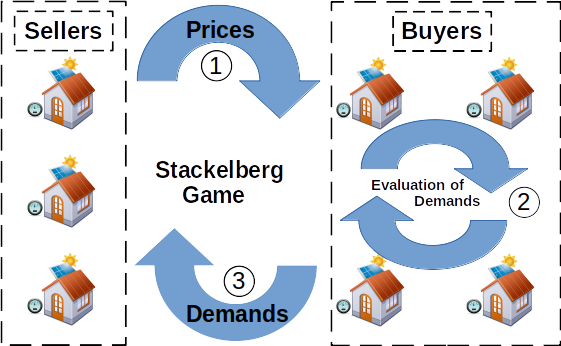}
\caption{System model and iterations.}
\label{fig: System Model}
\end{figure}

In our case (see Fig.~\ref{fig: System Model}), in the first iteration, sellers propose selling prices for their excess electricity \circled{1}. Buyers calculate the total demand for each seller in accordance with the prices offered~\circled{2}. Calculated demands for each seller are sent back to sellers~\circled{3}, and sellers update their prices which respect demands. Further iterations are performed on the same loop until the equilibrium point is reached.

\subsection{Thread Model and Assumptions}

    Buyers and sellers are honest-but-curious. 
    They follow protocol specifications, but may try to learn individual sellers' or buyers' sensitive data.
    External entities are not trustworthy. They may try to eavesdrop data in transit or intercept and alter the data.
  We assume that the entities communicate over secure and authentic communication channels.
   

\subsection{Privacy Requirements}
\label{PrivacyRequirements}
\begin{itemize}
    
    \item \textit{Seller price confidentiality}: Seller prices should be hidden from buyers as prices can reveal sensitive data about electricity consumption/production patterns of sellers. 
    
    \item \textit{Buyer demand confidentiality}: The total demand of buyers from a seller should be calculated in a privacy preserving way such that buyers can calculate  total demands without revealing the individual demand information. 
    
    \item \textit{Buyer profile variables confidentiality:} Buyer-specific sensitive profile variables 
    should be hidden from sellers.
    
\end{itemize}

\section{Privacy Preserving Energy Trading } \label{sec:PFET}

In this section, we propose a novel energy trading algorithm based on a competitive game theoretical approach 
that preserves buyers' and sellers' privacy by deploying HE scheme. 

    

\subsection{Game Theoretical Energy Trading Algorithm} \label{trading_algorithm}

We propose a Stackelberg Game in line with Paudel et. al.~\cite{paudel2018peer} in such a way that HE scheme can be deployed. 
Table~\ref{table:Notations} lists the notations used in the paper.

\begin{table}[!t]
 \footnotesize
\caption{Notations.}
\label{table:Notations}
\centering
 \footnotesize
\begin{center}
 \begin{tabular}{c l} 
 \hline
 \textbf{Symbol} & \textbf{Meaning} \\
 \hline
 
 $s_j$,  $b_i$  & $j$-th seller, $i$-th buyer \\
 $N_S$, $N_B$  & Number of sellers, buyers  \\ 
 $t_k$ & $k^{th}$ trading period\\
 $\pi_j$,  $S_j$   & Price asked by $s_j$, Supply that $s_j$ can provide  \\
 $D_j$, $\gamma_j$  & Total demand on $s_j$, State of $s_j$  \\
 $\lambda_i$, $\theta_i$ & Profile variables of $b_i$, linked to prosumers' behaviours.\\
 ${U_i}$ & Utility function of $b_i$\\
 ${W_{B_J}}$ & Welfare, utility of all buyers obtained from $s_j$ \\
  $\overline{W}$ & Average welfare \\
 $X_{ji}$  & Amount of electricity that $b_i$ wishes to buy from $s_j$ \\
 $PK_s$, $SK_s$  & Public, Private key pair of sellers  \\
 $Evk$ & Evaluation Key\\
 \hline
\end{tabular}
\end{center}
\end{table}

In the proposed Stackelberg Game, the main purpose of sellers is to maximise their revenues in a non-cooperative and competitive way while buyers aim to maximise their utilities when they buy certain amount of energy from the sellers. The system is initialised by the sellers (see Algorithm \ref{algo1}). For each trading period, $t_k$, new prices {[$\pi_1, ..., \pi_{N_S}$]} are offered by the sellers, e.g. $s_j$. The prices are sent to buyers and, as a response, demands proposed by the buyers for each seller $[D_1, ..., D_{N_S}]$ are returned back to the sellers after evaluations had been performed by buyer's algorithm (Algorithm \ref{algo2}). With the demand values returned, new prices are updated for each $s_j$ by the difference between demand for $s_j$ and maximum amount of electricity that $s_j$ can supply $S_j$. If the demand is larger than supply, the price is increased, otherwise -- reduced. The difference is multiplied by a small constant $\eta_1$ to prevent fluctuations on price updates and to provide better convergence to equilibrium. Updated  prices are bounded by $ \rho_{sell} < \pi_j^{t+1} < \rho_{buy} $
where $\rho_{sell}$ is FiT and $\rho_{buy}$ is retail price. Sellers' algorithm is run until demand and supply match.

\begin{algorithm}
  \caption{Sellers' Algorithm}
  \label{algo1}
   \small
   \SetKwInOut{Input}{Input}
    
    \Input{Number of Sellers, Buyers [$N_S$, $N_B$]}
    \
    initialization\;
    \For{Time t}{
        Propose Prices [$\pi_1, \pi_2, \pi_3, ..., \pi_{N_S}$]\;
        \Do{$|D_j - S_j| > \epsilon$ \textit{[For Each Seller j]}}{
            $[D_1, D_2, ..., D_{N_S}] \leftarrow Buyer's Algorithm( Prices )$\;
            
            \For{Each seller j}{
                $\pi_j^{t+1}\leftarrow \pi_j^{t} + \eta_1 \times (D_j^t - S_j)  $
                \;
                $\pi_j^{t+1} \leftarrow min(\rho_{buy},max(\rho_{sell}, \pi_j^{t+1}))$
                \;
            }
        }
   }

\end{algorithm}

The utility function \eqref{utility} of prosumers which is used to quantify the level of satisfaction that the prosumer have when it consumes certain amount of energy, is defined as
\begin{equation}
u(x_n) = \lambda_n \times x_n - {\theta_n \over 2} \times {x_n}^2 
\label{utility}
\end{equation}
where $x_n$ is the amount of energy consumed by prosumer n. $\lambda_n$ and $\theta_n$ are the profile variables of $b_n$, characterising prosumers' behaviours. 

The net utility \eqref{netutility}, $U_i$ of a buyer, $b_i$, when it buys certain amount of energy from seller, $s_j$  is obtained after subtraction of cost of the energy, from the utility function. 

\begin{equation}
U_i = u(x_{i}) - \pi_j \times x_{i}
\label{netutility}
\end{equation}

In Buyer's algorithm (Algorithm \ref{algo2}), first of all, the amount of energy that a buyer, $b_i$, wishes to buy from $s_j$, $X_{ji}$, is calculated. The aim of the buyer, $b_i$ is to maximise the utility function, $U_i$ in \eqref{netutility} w.r.t. price $\pi_j$ offered by the seller, $s_j$. The equation in line 4 of Algorithm \ref{algo2} which is used to calculate $X_{ji}$ maximising the $U_i$ is obtained after taking derivative of \eqref{netutility} and equating it to zero. After receiving the price $\pi_j$ offered by seller, $s_j$, each buyer, $b_j$ calculates $X_{ji}$ w.r.t. $\pi_j$ in ln. 4 of Algorithm \ref{algo2}. $X_{ji}$ is inversely proportional to price offered by $s_j$. When the price is high, buyers wish to buy less energy and vice versa.  The minimum value for $\lambda_i$ should be higher than maximum retail price. However, it is in the buyer's interest to offer $\lambda_i$ as low as possible, setting $\lambda_i$ close to retail price.  $\lambda_i$ is given a constant upper limit to guarantee convergence.

 The Welfare function \eqref{welfare} of buyers,  ${W_{B_J}}$ 
 is defined as the accumulated utilities of all buyers obtained when they buy electricity from $s_j$. $W_{B_j}$ is calculated in ln. 6 of Alg. 2.
 \begin{equation}
   W_{B_j} = \sum^{b_i}{U_i} = {1 \over 2} \sum^{b_i} {\theta_i  X_{ji}^2}
   \label{welfare}
 \end{equation}
 
\begin{algorithm}[!h]

 \caption{Buyers' Algorithm}
 \label{algo2}
\small
\SetAlgoLined
    Global Equilibrium States [$\gamma_1, \gamma_2, \gamma_3, ..., \gamma_{N_S}$]\;
    \SetKwInOut{Input}{Input}
    \SetKwInOut{Output}{Output}
    \Input{Prices [$\pi_1, \pi_2, \pi_3, ..., \pi_{N_S}$]}
    
    \Output{Demands [$D_1, D_2, D_3, ..., D_{N_S}$]}

    \For{Each Seller j}{
        \For{Each Buyer i}{
            $X_{ji}= {(\lambda_i - \pi_j)}/\theta_i$\;  
        }

        ${W_{B_J}} \leftarrow  {1 \over 2} \times \sum^{b_i} {\theta_i \times X_{ji}^2}  $ \;
    }
    $\overline{W} \leftarrow  \sum^{s_j} {\gamma_j \times W_{B_J}}  $ \; 
    
    \For{Each seller j}{
        
        $ D_j^{t} \leftarrow  \gamma_j^{t} \times \sum^{b_i} { X_{ji}}  $ \;  
        
        $\gamma_j^{t+1} \leftarrow \gamma_j^{t} + \eta_2 \times \gamma_j^{t}  \times (W_{B_J} - \overline{W}) $ \;

    }

\end{algorithm}

State $\gamma_j^{t}$ is the probability of $b_i$ choosing $s_j$ at time $t_k$. In the first trading period  $t_1$, Global Equilibrium States  [$\gamma_1, \ldots, \gamma_{N_S}$] are initialized with equal probabilities,  \textit{i.e. $\gamma_j=1/N_S$}. In the following trading periods, latest calculated states are used until the algorithm reaches an equilibrium point.

 Average welfare is calculated in line 8 as the accumulation of welfare of buyers multiplied by states. Total amount of energy buyers wish to buy from $s_j$ is multiplied by probability of seller $\gamma_j^{t}$ to calculate demand on $s_j$ in line 10. Finally, states are updated in line 11. If the welfare of buyers from $s_j$, ${W_{B_J}}$, is higher than the average welfare $\overline{W} $, then the probability of $s_j$ being selected, $\gamma_1 $, is increased; otherwise -- decreased by the difference between ${W_{B_J}}$ and $\overline{W} $. The difference is multiplied by a small constant $\eta_2 $ to avoid fluctuations on states.

\subsection{{Energy Trading Algorithm with Homomorphic Encryption  } } \label{privacy_techniques}






We propose a privacy-preserving version of the Game Theoretical Energy Trading Algorithm, which uses Fully Homomorphic Encryption (FHE) scheme. HE allows users to perform computations on encrypted data without revealing the plain data to anyone.  Results obtained after computations, when decrypted, is an identical output to that produced without using any encryption scheme. We deploy FHE which permits both addition and multiplication on encrypted data. There are other options which can provide privacy for the proposed system which are Differential Privacy and MPC. Differential privacy adds noise to input data to provide privacy in which accuracy of the data is lost in some degree. As we need accurate output data for both prices and demands, we have not considered differential privacy for our proposed trading platform. In MPC, entities jointly compute a function over their inputs without revealing actual data. However, the method has high communication intensity, and due to this, MPC has not been considered for our proposed platform having iterative communication among the entities.  

We represent the arithmetic operations on encrypted data using FHE as follows: $Eval(Evk,f,c_1,....c_n)$, where $Evk$ is an evaluation key, $f$ is an arithmetic operation and $c_1,....c_n$ are variables. The operation $f$ can be either multiplication (MULT) or addition (ADD). Input variables can be either encrypted or non-encrypted data. 



\begin{figure}[t] 
\includegraphics[scale=0.31]{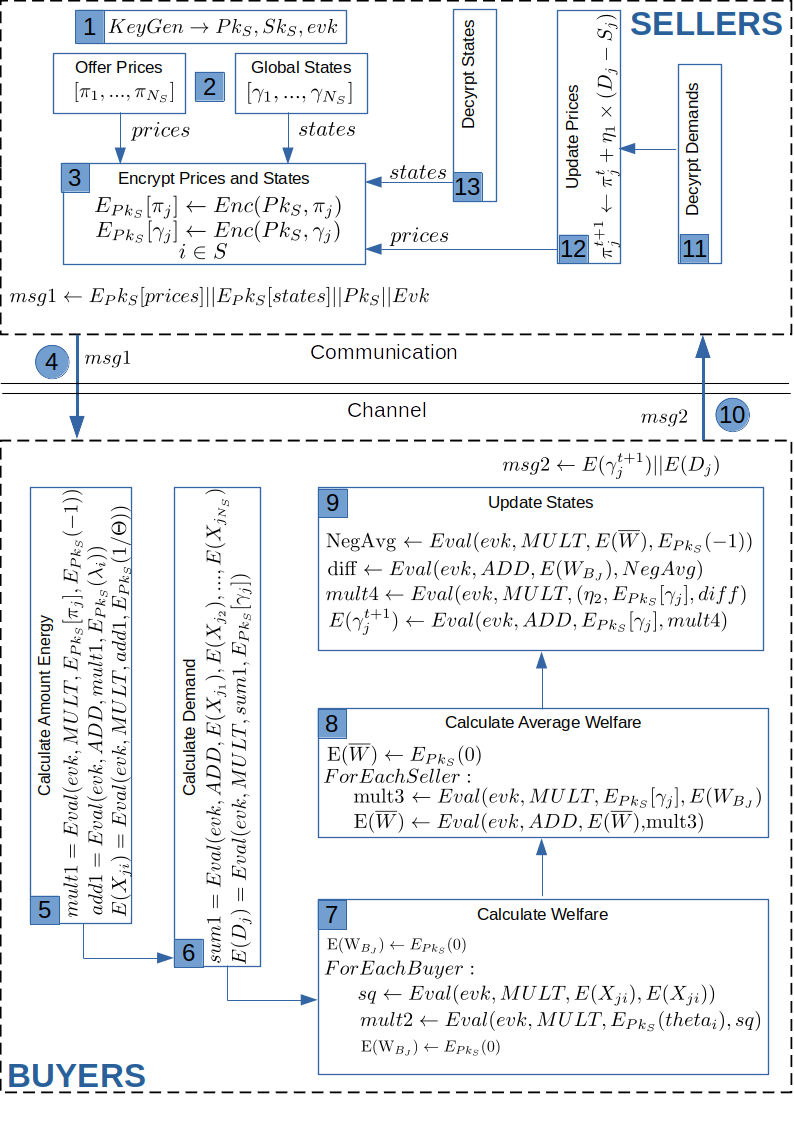}
\caption{Privacy-friendly Energy Trading platform.}
\label{fig: PET}
\end{figure}

As depicted in Fig.~\ref{fig: PET}, first of all, public/private key pair of sellers, $PK_s$/$SK_s$, and Evaluation Key, $Evk$, are generated in \squaredNum{1}. After this, trading period starts and prices are offered by the sellers in \squaredNum{2}. Global Equilibrium States [$\gamma_1, ..., \gamma_{N_S}$] are also initialised  in \squaredNum{2}. Initialised prices and states are encrypted using $PK_s$ in \squaredNum{3}. $E_{{PK}_{S}}[prices], E_{{PK}_{S}}[states], PK_S, Evk $ are concatenated to $msg1$ and sent to buyers in \circled{4}. The encrypted amount of energy that $b_i$ wishes to buy from $s_j$, $E(X_{ji})$, is calculated in \squaredNum{5}. Before the calculation, buyer specific variables $\lambda_i$ and $\theta$ are encrypted using the $PK_s$ received from sellers. $E(X_{ji})$ is calculated for each $s_j$ and $b_i$ combination, so block \squaredNum{5} is run for  $N_S$ x $N_B$ times. Encrypted demands $E(D_j)$ are calculated in \squaredNum{6} using $E(X_{j_2}),...,E(X_{j_{N_S}})$ and $E(X_{j_{N_S}})$ for each seller, so block \squaredNum{6} runs for $N_S$ times.

After this, encrypted welfare of buyers when they buy electricity from $s_j$, $E({W_{B_j}})$s are calculated over previously encrypted variables in \squaredNum{7} for each $s_j$, so it is run $N_S$ times. Encrypted average welfare $E(\overline{W})$ is calculated using the previously calculated $E({W_{B_j}})$s in \squaredNum{8}. States are updated $E(\gamma_j^{t+1})$ in encrypted format using $E(\overline{W})$ and $E({W_{B_j}})$ in \squaredNum{9} for each seller, which means block \squaredNum{9} is run $N_S$ times. Updated states in \squaredNum{9} and Encrypted demands in \squaredNum{6} are concatenated and sent to the seller side in encrypted format in \circled{10}. 

After demands has been decyrpted in \squaredNum{11}, new prices to be offered by the sellers for the next iterations is calculated in \squaredNum{12} in non-encrypted format by the difference between demand $D_j$ and $S_j$. If the difference is low, such that $|D_j - S_j| < \epsilon$, the trading period terminates. Otherwise, updated states $E(\gamma_j^{t+1})$ obtained from {\circled{10}}  are decrypted in \squaredNum{13} and forwarded along with updated prices $\pi_j^{t+1}$ into \squaredNum{3} to be used in the next iterations. The reason to send updated states from buyers to sellers and decrypt, encrypt them on the seller side and return them to buyers is to eliminate noise in encrypted states.

\begin{figure*}
  \centering
  \begin{subfigure}{.33\linewidth}
    \centering
    \includegraphics[width = \linewidth]{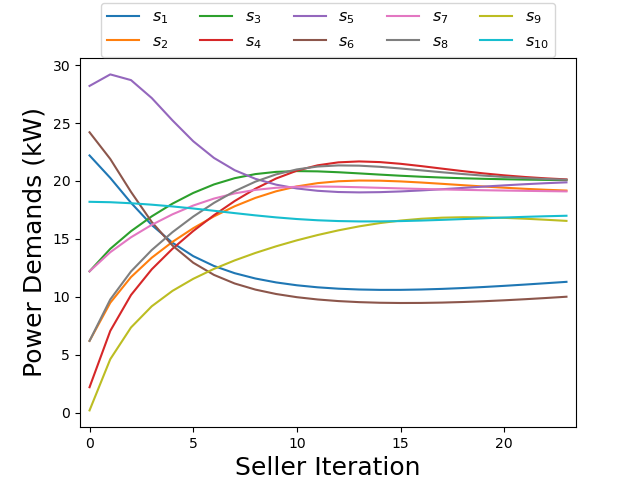}
   
  \end{subfigure}%
  \hspace{0.01em}
  \begin{subfigure}{.33\linewidth}
    \centering
    \includegraphics[width = \linewidth]{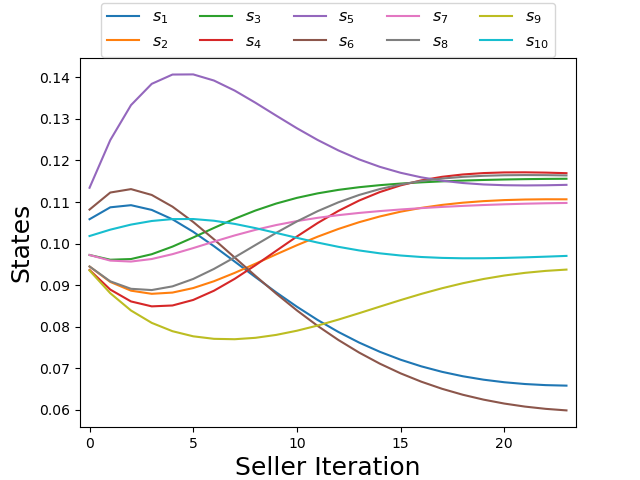}
   
  \end{subfigure}%
  \hspace{0.01em}
  \begin{subfigure}{.33\linewidth}
    \centering
    \includegraphics[width = \linewidth]{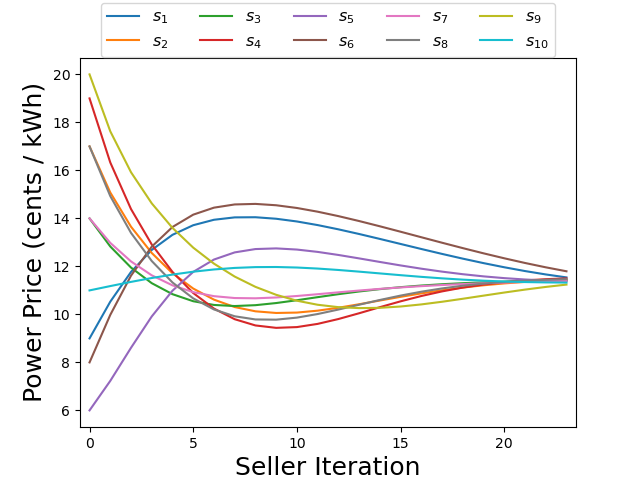}
    
  \end{subfigure}
  \caption{Simulation Results.}
  \label{fig:SimulationResults}
\end{figure*}

\section{{Equilibrium and Time Complexity Analysis}}
\label{equilibrium_time_complexity}


\textbf{Equilibrium analysis:}
Nash equilibrium (NE) exists in the game if the following conditions are satisfied{~\cite{wang2014game}}: 1) player set is finite, 2) strategy sets are bounded, convex and closed and 3) utility functions are continuous and  concave in the strategy space.
In our design, number of players which are comprised of sellers and buyers in the player set is finite.

Sellers determine their strategies with prices they propose and buyers determine their strategies with the amount of energy that a buyer, $b_i$, wishes to buy from $s_j$, $X_{ji}$. The prices have upper and lower limit such that $ \rho_{sell} < \pi_j < \rho_{buy} $. $X_{ji}$ depends on $\pi_j$, $\lambda_i$ and $\theta_i$ which are either limited with upper and lower bounds, or constants, which means $X_{ji}$ has upper and lower limits. Therefore, strategy sets comprised of $\pi_j$ and $X_{ji}$ are non-empty, closed, bounded and convex.

The aim of the buyers is to maximise $U_i$ with strategies determined by $X_{ji}$. Second derivative of $U_i$ w.r.t. $X_{ji}$ is given,

\begin{equation}
    \frac{\partial^2 U_i}{\partial x_{ji}^2} = - \theta_i < 0
\end{equation}

so, $U_i$ is strictly concave in $X_{ji}$. Therefore, we can conclude that there exists a NE. {The analysis is valid for both algorithms with HE and without HE as the same operations are performed with same data in non-encrypted or encrypted format.    }

\textbf{Time complexity analysis:}
 In the Game Theoretical Energy Trading algorithm proposed, Sellers' Algorithm which act as a leader of Stackelberg Game is the main function in which Buyers' Algorithm is called. Hence, time complexity is related to the input of Sellers' Algorithm: number of sellers and buyers [$N_S$, $N_B$]. Double loop starting from ln. 2 in Algorithm \ref{algo2} has a quadratic $O(n^2)$ time complexity. Ln. 10 in Algorithm \ref{algo2} is quadratic due to loop and sum operation. Other parts have either O(n) or O(1) complexity. As a result, time complexity of the Game Theoretical Energy Trading algorithm is quadratic: $O(n^2)$. {Time complexity of algorithm with HE is $O(n^2)$  too, as the blocks having highest time complexity in Fig.~\ref{fig: PET} are run for  $N_S$ x $N_B$ times. }

\section{{Security and Privacy Analysis}} \label{security_analysis}

\textit{Seller price confidentiality}: Seller prices are encrypted with the public key, $PK_S$, before being sent to buyers. As sellers are the only ones who have access to the corresponding private key, $SK_S$, buyers cannot see the prices of sellers.

\textit{Buyer demand confidentiality}: Total buyer demand per seller is calculated on encrypted data which can only be decrypted using the private key of sellers, hence only sellers can access to these total demands. 

\textit{Buyer profile variables confidentiality:} Although, buyer  profile  variables are encrypted with $PK_S$ to be used for calculations on the buyer side, only the result of computations are sent to the sellers, not the individual buyer profile  variables. Hence, sellers cannot trace back the buyer profile  variables.

\begin{table}[t]
\caption{Computational cost for PFET per iteration.}
\label{table:computationCost}
\centering
\footnotesize

\begin{center}
 \begin{tabular}{l | c  c c c c  } 
 \hline



\textbf{Number of sellers, $N_S$}      & 10      & 20    & 30    & 40    & 50 \\
\textbf{Number of buyers, $N_B$}      & 10      & 20    & 30    & 40    & 50 \\
\hline
\textbf{Time Spent} & $7.8s$  & $20.7 s$ & $39.3s$ & $62.5s$ &  $89.7s$ \\

 
 



 \hline
\end{tabular}
\end{center}
\end{table}

\section{{Simulation Results}} \label{performance_evaluation}

We implemented PFET using Python 3.8.5 programming language. Pyfhel library~\cite{Pyfhel} is used for FHE operations, which uses Microsoft SEAL library~\cite{SEAL} as a back-end. We run simulations on a Laptop with the following parameters: CPU -- Intel(R) Core(TM) i5-8350U CPU @ 1.70GHz and System Memory -- 8GB. The following parameters are set for the simulations: $\pi_{min} = 4$ cent (FiT), $\pi_{max} = 20$ cent (Retail Price), $\eta_1 = 0.15$ , $\eta_2 = 0.0001$ and $\lambda_i = 20.1$, $\theta_i = 0.5$. 

{ Number of iterations to reach equilibrium depends on initial supply values and prices offered by the sellers. When initial supply values and/or prices are correlated with each other, the number of iterations decrease. As an example, when the initial offered prices of all sellers are replaced with 5 cents/kWh, the number of iterations reduces to 10 for the setup in Fig.~\ref{fig:SimulationResults}. Also, the number of iterations can be adjusted with $\eta_1$ such that when $\eta_1$ is set to higher values, number of iterations decrease. $\eta_1$ is set to `0.15' in accordance to have low number of iterations but also not cause fluctuations for the proposed setup. $\eta_2$ is set to a small number `0.0001' to fit high order of welfare values to low order of states. It is in the buyer's interest to offer $\lambda_i$ as low as possible, setting $\lambda_i$ close to retail price, so $\lambda_i$ is set to `20.1'. Buyer specific parameter  $\theta_i $ is set to the same value 0.5 for each buyer for the sake of clarity.}

 We vary the total number of sellers and buyers from $20$ to $100$ and measure the total time our proposed solution takes with and without HE in place. Average computational costs {for the energy trading algorithm with HE} per each iteration are presented in Table~\ref{table:computationCost}. Average execution time per iteration with HE almost fit to the equation $2.85n^2 +4.35n +0.1$  where $n$ is correlated to the number of users (i.e. $n = (N_B+N_S)/20$). This equation also confirms that time complexity of the algorithm is $O(n^2)$. The computation times of the algorithm without HE are around  $ \approx 0.02s$. 

Simulation outputs of an example case with energy supplies of sellers $S = [12, 19, 20, 20, 20, 11, 19, 20, 16, 17]~kW$ and with initial prices $\pi =  [9, 17, 14, 19, 6, 8, 14, 17, 20, 11]~cent$ offered by sellers, are illustrated in Fig.~\ref{fig:SimulationResults}.  Buyers' demands reach to equilibrium points and \textit{match} energy supplies of the sellers. States values reach to equilibrium in the same way buyers' demands do. Prices starting from initially proposed values also converge to an equilibrium point where final prices are between FiTs and retail price.


\section{{Conclusion} } 

\label{conlcusion_future_work}


We proposed a privacy preserving \textit{competitive} energy trading platform, PFET, based on game theoretical approach and fully homomorphic encryption scheme. The evaluation results confirm that PFET incentives both buyers and sellers in terms of prices, demands and supplies match when the system reaches an equilibrium point. Sellers' sensitive data is protected from buyers and vice versa. 
As a future work, we plan to extend PFET by (i) considering network costs and fees and (ii) investigating the trade off between utility-privacy-performance when deploying diffrenet types of HE schemes.  

\bibliographystyle{IEEEtran}

\bibliography{sample.bib}

\end{document}